\documentclass[runningheads]{llncs}
\usepackage[utf8]{inputenc}
\usepackage[noadjust]{cite}
\usepackage{epsfig}
\usepackage{epstopdf}
\usepackage{graphicx}
\usepackage{amsfonts}
\usepackage{amssymb}
\usepackage{amsmath}
\usepackage{color}
\usepackage{comment}
\usepackage{bm}

\newcommand{\adjacent}{\textsf{adjacent}}
\newcommand{\degree}{\textsf{degree}}
\newcommand{\neighbor}{\textsf{neighborhood}}
\newcommand{\multiplicity}{\textsf{multiplicity}}

\newcommand{\Order}{{\rm O}}
\newcommand{\order}{{\rm o}}
\begin{document}
\title{Succinct Data Structures for Series-Parallel, Block-Cactus and $3$-Leaf Power Graphs}
\titlerunning{Succinct Data Structures for SP, Block-Cactus and $3$-Leaf Power Graphs}
%
\author{Sankardeep Chakraborty\inst{1}\orcidID{0000-0002-2395-4160} \and
Seungbum Jo\inst{2}\orcidID{0000-0002-8644-3691} \and
Kunihiko Sadakane\inst{3}\orcidID{0000-0002-8212-3682} \and
Srinivasa Rao Satti\inst{4}\orcidID{0000-0003-0636-9880}}
\authorrunning{Chakraborty, Jo, Sadakane and Satti}
%
\institute{ The University of Tokyo, Japan \\\email{sankardeep@mist.i.u-tokyo.ac.jp} \and
Chungnam National University, South Korea \\\email{sbcho1204@gmail.com} \and
The University of Tokyo, Japan \\\email{sada@mist.i.u-tokyo.ac.jp} \and
Norwegian University of Science and Technology, Norway \\\email{srinivasa.r.satti@ntnu.no}}
\maketitle              
\begin{abstract}
We design succinct encodings of {\it series-parallel, block-cactus} and {\it 3-leaf power} graphs while supporting the basic navigational queries such as degree, adjacency and neighborhood {\it optimally} in the RAM model with logarithmic word size. One salient feature of our representation is that it can achieve  optimal space even though the exact space lower bound for these graph classes is not known. For these graph classes, we provide succinct data structures with optimal query support for the first time in the literature. For series-parallel multigraphs, our work also extends the works of Uno et al. (Disc. Math. Alg. and Appl., 2013) and Blelloch and Farzan (CPM, 2010) to produce optimal bounds. 
\end{abstract}

\setcounter{footnote}{0}

\section{Introduction}
In modern algorithm development, we observe two drastically opposing trends. Even though memory capacities are increasing and their prices are drastically reducing day-by-day, input data sizes that are being stored are growing at a much faster pace, and this is due to the ongoing digital transformation of business and society in general. There are many application areas, e.g., social networks, web mining, and video streaming systems, where already there exists a tremendous amount of data and it is only increasing. 
In these domains, most often, a natural representation of the underlying data sets is in the form of graphs, and with each passing day, these graphs are becoming massive. To process such huge graphs and extract useful information from them, we need to answer the following two concrete questions among others: (1) can we store these massive graphs in compressed form using the minimum amount of space? and (2) can we build space-efficient indexes for these huge graphs so that we can extract useful information about them by executing efficient query algorithms on the index itself? The field of {\it succinct data structures} aims to exactly answer these questions satisfactorily, and it has been one of the key contributions to the algorithm community in the past two decades, both theoretically and practically. More specifically, given a class of certain combinatorial objects, say $T$, from a universe $U$, the main objective here is to store any arbitrary member $x \in T$ using the \textit{information-theoretic lower bound} of $\log(|U|)$ bits (in addition to $o(\log(|U|))$ bits) \footnote{Throughout the paper, we use logarithm to the base 2.} along with efficient support of a relevant set of operations on $x$. 

There exists already a large body of work representing various combinatorial structures succinctly along with fast query support. For example, succinct data structures for rooted ordered trees~\cite{Jacobson, MunroR01,NavarroS14,Raman013}, chordal graphs~\cite{MunroW18}, graphs with treewidth at most $k$~\cite{FarzanK14}, separable graphs~\cite{BlellochF10}, interval graphs~\cite{DBLP:conf/wads/AcanCJS19} etc., are some examples of these data structures. Following similar trend, in this work we provide succinct data structures
for series-parallel multigraphs~\cite{doi:10.1002/sapm194221183}, block-cactus graphs~\cite{GurevichSV84} and $3$-leaf power graphs~\cite{BrandstadtL06}. We defer the definitions of the graph classes to the individual sections where their succinct data structure is proposed. These graphs are important because not only they are theoretically appealing to study but they also show up in important practical application domains, e.g., series-parallel graphs are used to model electrical networks, cacti are useful in computational biology etc. 
To the best of our knowledge, our work provides succinct data structures with optimal query support for the first time in the literature (although there exists  a succinct data structure for simple series-parallel graphs~\cite{BlellochF10}, such a structure is not known for series-parallel multigraphs). 

\subsection{Previous work}\label{subsec:previous}

\noindent
{\bf Series-Parallel (SP) graphs.}
The information-theoretic lower bound (ITLB) for encoding a simple SP-graph with $n$ vertices is $3.18n+o(n)$ bits~\cite{DBLP:journals/ejc/BodirskyGKN07} whereas the ITLB for encoding an  SP multigraph with $m$ edges is $1.84m + o(m)$ bits~\cite{doi:10.1142/S179383091360001X}.
Since an SP graph is {\em separable}, one can obtain a succinct representation of any SP graph by using the result of Blelloch and Farzan~\cite{BlellochF10} while supporting some navigation queries efficiently. However, this only works for simple SP graphs~\cite{DBLP:reference/algo/MunroN16} since one cannot store the look-up table for all possible micro-graphs (containing multi SP graphs with any fixed number of vertices) within the limited space (as the number of edges is not bounded)\footnote{Note that one can encode SP multigraphs by encoding the underlying simple graph using Blelloch and Farzan's encoding, along with a bit string of size $m$ to represent the multiplicities of the edges. However, the space usage is not succinct in this case.}. 
Also since simple SP graphs are exactly the class of graphs with treewidth $2$, one can use the data structure of Farzan and Kamali~\cite{FarzanK14} for representing SP graphs but again, this only works for simple SP graphs. 
For multigraph case with $m$ edges, Uno et al.~\cite{doi:10.1142/S179383091360001X} present an encoding for SP multigraphs taking at most $2.53 m$ bits without supporting any navigational queries efficiently. 

\noindent
{\bf Block-Cactus and 3-Leaf Power graphs.}
The ITLB for encoding a block-cactus graph and a $3$-leaf power graph with $n$ vertices are $2.092n + o(n)$~\cite{block-cactuslb} and $1.35n+o(n)$~\cite{DBLP:conf/analco/ChauveFL17} bits respectively. Note that the class of Block-cactus graphs contains both \textit{cactus} and \textit{block} graph classes. As any cactus graph is planar, and hence separable, one can again use the result of Blelloch and Farzan~\cite{BlellochF10} to encode them optimally with supporting the navigation queries efficiently. However, this approach doesn't work for block or block-cactus graphs since they are not separable. 


\subsection{Our Main Contribution}

We design succinct data structures for (i) series-parallel multigraphs in Section~\ref{sec:spgraph} and (ii) block-cactus graphs in Section~\ref{sec:block},
and finally (iii) 3-leaf power graphs in Section~\ref{sec:3leaf}
to support the following queries.
Given a graph $G=(V,E)$ and two vertices $u,v\in V$, (i) $\degree{}(v)$ returns the number of edges incident to $v$ in $G$, (ii) $\adjacent{}(u,v)$ returns true if $u$ and $v$ are adjacent in $G$, and false otherwise, and finally (iii) $\neighbor{}(v)$ returns the set of all (distinct) vertices that are adjacent to $v$ in $G$. 
The following theorem summarizes our main results on these graphs.

\begin{theorem}\label{thm:all}
There exists a succinct data structure that supports $\degree{}(u)$ and $\adjacent{}(u, v)$ queries in $\Order(1)$ time, and $\neighbor(u)$ query in $\Order(\degree{}(u))$ time, 
for (1) series-parallel multigraphs, 
(2) block-cactus graphs, 
and (3) $3$-leaf power graphs.
\end{theorem}

The reason for considering these three (seemingly unrelated) graph classes is that any graph in each of these three classes has a corresponding tree-based representation - and hence these graphs can be encoded succinctly by encoding the corresponding tree. 
In what follows, we briefly discuss a high level idea on how to succinctly represent the graphs of our interest. Roughly speaking, given a graph $G$ ($G$ could be series-parallel, block-cactus or $3$-leaf power), we first convert it to a labeled tree $T_G$ which can be used to decode $G$. We then represent $G$ by encoding $T_G$ using the \textit{tree covering  (TC) algorithm} of Farzan and Munro~\cite{FarzanM13}, which supports various tree navigation queries in $\Order(1)$ time.
However, we cannot obtain directly the succinct representation of $G$ with efficient navigation queries from the tree covering of $T_G$. 
More specifically, the tree covering algorithm first decomposes the input tree and encodes each decomposed tree separately. Thus, a lot of information of $G$ can be lost in each of the decomposed trees. 
For example, decomposed trees may not even belong to the graph class that we originally started with in the first place (and this is in stark contrast to the situation while designing succinct data structures for trees). Thus, we need to apply non-trivial local changes (catering to each graph class separately) to these decomposed trees and argue that (i) these changes convert them again back to the original graph class, without consuming too much space, and (ii) navigation queries on $G$ can be supported  efficiently as tree queries on $T_G$.
As a consequence, one salient feature of our approach is that for the graphs $G$ we consider in this paper, it is not necessary to know the exact information-theoretic lower bound, to design succinct data structures for them if we only know the asymptotic lower bound of the number of non-isomorphic graphs of $G$ with a given number of vertices. 
Note that the overall idea of `encoding the graph as a tree-based representation and using the TC algorithm to encode the tree to support the navigation operations on the graph' is subsequently used in~\cite{dcc-clique} to obtain succinct representation for graphs of small clique-width. The other main contribution of this paper is to construct suitable tree-based encodings and showing how to adapt the TC representation to support the operations.


\section{Preliminaries and Main Techniques}\label{sec:prelim}


Throughout our paper, we assume familiarity with succinct/compact data structures (as given in~\cite{Navarro}), basic graph theoretic terminology (as given in~\cite{Diestel}), and graph algorithms (as given in~\cite{CLRS}). All the graphs in our paper are assumed to be connected and unlabeled, i.e., we can number the vertices arbitrarily. 
Moreover, we assume the usual model of computation, namely a $\Theta (\log n)$-bit word RAM model where $n$ is the size of the input 
($n$ is the number of vertices in the case of graphs, and the number of edges in the case of multigraphs). We start by sketching a modification to the tree covering algorithm of Farzan and Munro~\cite{Farzan2014}.

\subsection{Tree covering}

The high level idea of the tree covering algorithm is to decompose the tree into subtrees called 
mini-trees (in the rest of the paper, we use {\em subtree} to denote any connected subgraph of a given tree), and further decompose the mini-trees into yet smaller subtrees called {\it micro-trees}. The micro-trees are small enough to be stored in a compact table. The root of a mini-tree can be shared by several other mini-trees. To represent 
the tree, we only have to represent the connections and links between the subtrees. In what follows, we summarize the main result of Farzan and Munro~\cite{Farzan2014} in the following theorem:


\begin{theorem}[\cite{Farzan2014}]\label{th:originalTC}
For a rooted ordered tree with $n$ nodes and a positive integer $1 \leq L \le n$,
we can obtain a tree covering satisfying
(1) each subtree contains at most $2L$ nodes,
(2) the number of subtrees is $\Order(n/L)$,
(3) each subtree has at most one outgoing edge, apart from those from the root of the subtree.
\end{theorem}
For each subtree after the decomposition of Theorem~\ref{th:originalTC}, 
the unique node that has an outgoing edge is called the \textit{boundary node}
of the subtree, and the edge is called the \textit{boundary edge} of the subtree.
The subtree may have multiple outgoing edges from its root node (in this case, we call it a \textit{shared root node}),
and those edges are called \textit{root boundary edges}.

To obtain a succinct representation, 
we first apply Theorem~\ref{th:originalTC} with $L = \log^2 n$, to obtain $\Order(n/\log^2 n)$ mini-trees (here and in the rest of the paper, we ignore all floors and ceilings which do not affect the main result).  
The tree obtained by contracting each mini-tree into a vertex is referred to as the \textit{tree over mini-trees}.
If more than one mini-tree shares a common root, we create a dummy node in the tree
and make the nodes corresponding to the mini-trees as children of the dummy node.
We also set the parent of the dummy node as the node corresponding to the parent mini-tree.
(See Figure~\ref{fig:sp1} for an example.)
This tree has $\Order(n/\log^2 n)$ vertices
and therefore can be represented in $\Order(n/\log n) = \order(n)$ bits using
a pointer-based representation.  Then, for each mini-tree, we again apply Theorem~\ref{th:originalTC}
with parameter $\ell = \frac{1}{4} \log n$ to obtain $\Order(n/\log n)$ micro-trees in total.
The tree obtained from each mini-tree by contracting each micro-tree into a node, and adding dummy nodes for micro-trees sharing a common root (as in the case of the tree over mini-trees)
is called the \textit{mini-tree over micro-trees}.  
Each mini-tree over micro-trees has
$\Order(L/\ell) = \Order(\log n)$ vertices, and can be represented
by $\Order(\log (L/\ell)) = \Order(\log\log n)$-bit pointers.
For each non-root boundary edge of a micro-tree $t$, we encode from which vertex of $t$ it comes out and the rank among all children of the vertex. One can encode the position where the boundary edge is inserted in $\Order(\log \ell)$ bits.
Note that in our modified tree decomposition, each node in the tree is in exactly one micro-tree.

For each micro-tree, we define its representative
as its root node if it is not shared with other micro-trees,
or the next node of the root node in preorder if it is shared.
Then we mark bits of the balanced parentheses representation~\cite{MunroR01} of the entire
tree corresponding to the representatives.
If we extract the marked bits, it forms a balanced parentheses (BP) and it represents the mini-tree over micro-trees.
The positions of marked bits are encoded in $\Order(n \log\log n/\log n)$ bits
because there are $\Order(n/\log n)$ marked bits in the BP representation
of $2n$ bits.
The BP representation is partitioned into $\Order(n/\log n)$ many
variable-length blocks, each of which is of length $\Order(\log n)$.
We can decode each block in constant time.

To support basic tree navigational operations
such as parent, $i$-th child, child rank, degree, lowest common ancestor (LCA),
level ancestor, depth, subtree size, leaf rank, etc. in constant time,
we use the data structure of~\cite{NavarroS14}.
Note that we slightly change the data structure because now each block is of variable length.
We need to store those lengths, but it is done by using the positions of the marked bits. 

The total space for all mini-trees over micro-trees
is $\Order(n/\ell \cdot \log \ell) = \Order(n \log\log n/\log n) = \order(n)$ bits. Finally, the micro-trees are stored as two-level pointers (storing the size, and an offset within all possible trees of that size) into a precomputed table that contains the representations of all possible micro-trees. The space for encoding all the micro-trees using this representation can be shown to be $2n+\order(n)$ bits.

\subsection{Graph Representation Using Tree covering}\label{sec:graph_to_tree}
This section describes the high-level idea to obtain succinct encodings for the graph classes that we consider.
Let $\cal{C}$ be one of the graph classes among series-parallel multigraphs, block-cactus graphs, and 3-leaf power graphs. Then the following properties hold. 
\begin{itemize}
\item the ITLB for representing any graph $G \in \cal C$ is $kn + o(kn)$ bits for some constant $k > 0$~\cite{doi:10.1142/S179383091360001X, block-cactuslb, DBLP:conf/analco/ChauveFL17}, where $n$ is the number of vertices (block-cactus, and 3-leaf power graphs) or edges (series-parallel multigraphs) in $G$. 
\item For any connected graph $G \in \cal C$, there exists a labeled tree $T_G$ of $\Order(n)$ nodes, such that $G$ can be uniquely decoded from $T_G$. 
\end{itemize}

By the above properties, one can represent any graph $G \in \cal C$ by encoding the tree covering of $T_G$ (with $L = \log ^2 n$ and $\ell = \frac{\log n}{2k}$). Unfortunately, tree covering on $T_G$ does not directly give a succinct encoding of $G$ since the number of all non-isomorphic graphs in $\cal C$ can be much smaller than the number of all non-isomorphic labeled trees of the same size (for example, multiple labeled trees can correspond to the same graph).
To solve this problem, we maintain a precomputed table of all non-isomorphic graphs in $\cal C$ of size at most $\ell$, along with their corresponding trees in \textit{canonical representation}. By representing each micro-tree as an index of the corresponding graph in the precomputed table, we can store all the micro-trees of $T_G$ in succinct space. 
If a micro-tree $t$ does not have a corresponding graph in $\cal C$ (i.e., there is no corresponding graph in the precomputed table), we first extend $t$ to $T_g$ where $g \in \cal C$ of size at most $\ell$ by adding some \textit{dummy nodes}, and encode $t$ as the index of $g$, along with the information about dummy nodes.  
Since we only add a small number of (at most $\Order(1)$) dummy nodes for each micro-tree, all the additional information can be stored within succinct space. In the following sections, we describe how to add such dummy nodes for series-parallel, block-cactus, and 3-leaf power graphs. 

Finally, for the case when $G \in \cal C$ is not connected, we extend the above idea as follows. 
We first encode all the connected components of $G$ separately, and encoding the sizes of the connected components using the encoding of \cite{DBLP:journals/algorithmica/El-ZeinLMRC17, DBLP:journals/rss/SumigawaS19} using at most $\Order(\sqrt{n})$ additional bits. This implies we can still encode $G$ in succinct space even if $G$ is not connected. In the rest of this paper, we assume that all the graphs are connected. 

\section{Series-Parallel Graphs}\label{sec:spgraph}

Series-parallel graphs~\cite{doi:10.1002/sapm194221183} (SP graphs in short) are undirected multi-graphs which are  defined recursively as follows.
\begin{itemize}
    \item A single edge is an SP graph.  We call its two end points as terminals.
    \item Given two SP graphs $G_1$ with terminals $s_1, t_1$ and $G_2$ with terminals $s_2, t_2$,
    \begin{itemize}
        \item their \emph{series composition}, the graph made by identifying $t_1 = s_2$, is an SP graph
    with terminals $s_1, t_2$; and
        \item their \emph{parallel composition},
    the graph made by identifying $s_1 = s_2$, and $t_1 = t_2$, is an SP graph with terminals $s_1$ and $t_1$.
    \end{itemize}
    
    
\end{itemize}

\begin{figure}[t]
\begin{center}
\includegraphics[clip, width=12cm]{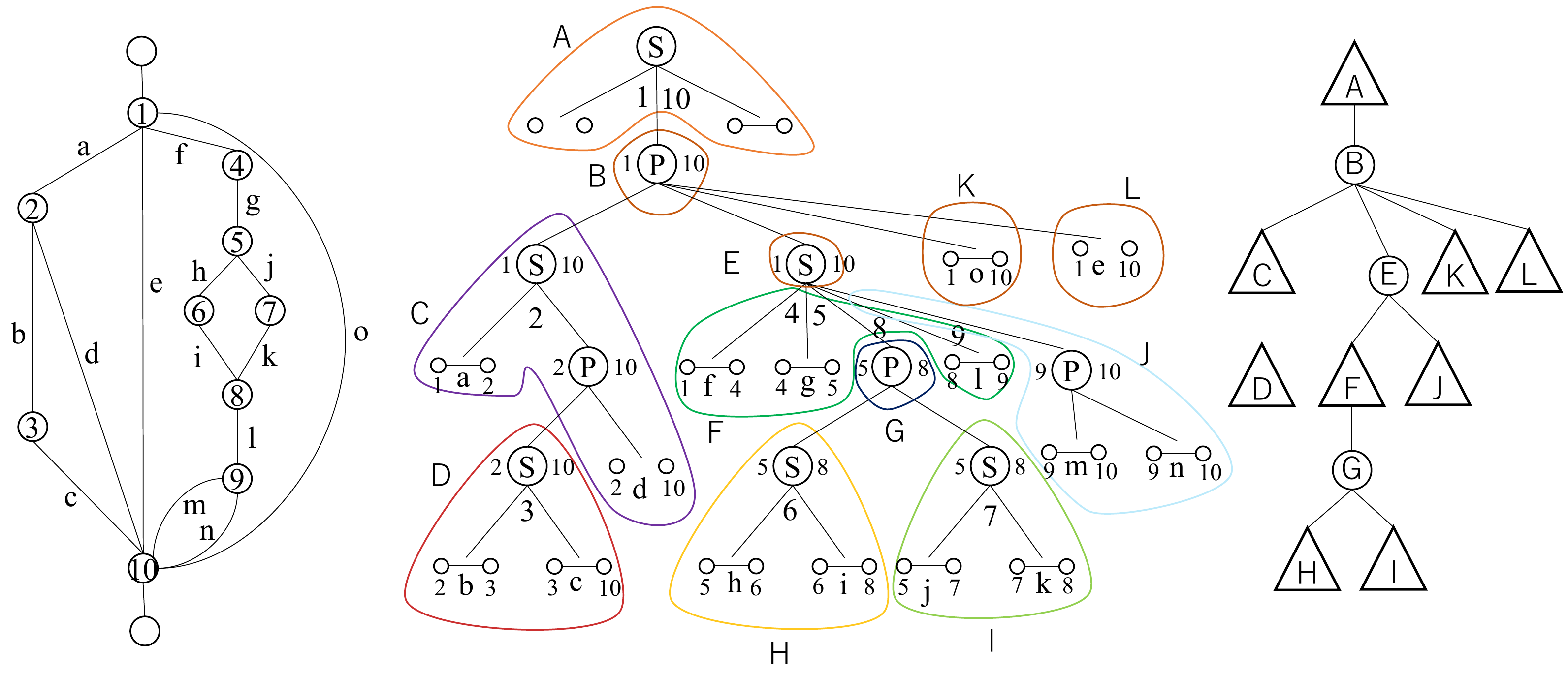}
\caption{Example of an SP graph (left), its SP tree representation with tree covering (middle), and
the tree over mini-trees (right).
The roots of mini-tree G and K are dummy nodes.
Numbers below S nodes are inorders.
Numbers besides internal nodes of the SP tree
are the left and right labels.
Leaves of the tree also have the left and right
labels, which are vertex labels
of the SP graph.
}
\label{fig:sp1}
\end{center}
\end{figure}

From this construction, we can obtain the binary tree $T$ representing an SP graph $G=(V, E)$ as follows.
Each leaf of the binary tree $T$ corresponds to an edge of $G$.
Each internal node $v$ of $T$ has a label S (or P), which represents an SP graph
made by the series (or parallel) composition of the two SP graphs represented by the two child subtrees of $v$.
We convert it into a multi-ary SP tree $T_G$
by merging vertically consecutive nodes with identical labels into a single node. More precisely, while scanning all the nodes in $T_G$ in bottom-up, we contract every edge $(v, v')$ if $v$ and $v'$ have the same labels.
Then all the internal nodes at the same depth have the same labels, and the labels alternate between the levels.
See Figure~\ref{fig:sp1} for an example.
Note that any two non-isomorphic SP graphs have different SP trees. \\

\noindent
\textbf{Succinct representation.}
Let $n$ and $m$ be the number of vertices and edges of $G$, respectively. Then $T_G$ has $m$ leaves, and $\Order(m)$ nodes. 
First, we construct the SP tree $T_G$ from an SP graph $G=(V, E)$.
If the root of $T_G$ is a P node, we add a dummy parent $r$ labeled S with three children, and make the original root as the middle child of $r$.
The first and the last children of $r$ correspond to dummy edges.
If the root of $T_G$ is an S node,
we also add two leaves as the leftmost and rightmost children of the root, corresponding to dummy edges.
We refer to this modified tree as $T_G$.
Let $s = \Order(m)$ be the number of nodes in $T_G$.
Then we apply the tree covering algorithm with parameters $L = \log^2 s$ 
and 
$\ell = (\log s)/4$.

It is obvious that each micro-tree without dummy leaf nodes represents an SP graph.
For each graph corresponding to a micro-tree, we use a linear time algorithm~\cite{DBLP:journals/siamcomp/ValdesTL82} to obtain a canonical representation of the micro-tree. Note that if the graphs corresponding to two micro-trees are isomorphic, then those two micro-trees have the same canonical representation.
We create a table to store all non-isomorphic SP graphs with at most $\ell$ vertices, and encode each micro-tree as a pointer into this table.
To reconstruct the original graph from the graphs corresponding to the micro-trees, we need additional information to combine these graphs.
More specifically, assume an SP graph $G$ consists of a series composition of graphs $G_1$ and $G_2$,
whose terminals are $s_1, t_1$, and $s_2, t_2$ respectively.
Then one can construct two different graphs $G$ and $G'$ by (i) connecting $t_1$ and $t_2$ or (ii) $t_1$ and $s_2$.
Thus, for each micro tree, we add one extra bit to store this information.

For each S node of $T_G$, we assign an {\em inorder number}~\cite{Sada07a} (we only assign inorder numbers for S nodes).
Inorder numbers in a rooted tree are given during a preorder traversal from the root.  
If a node $v$ is visited from one of its children
and another child of $v$ is visited next, we assign one inorder number to $v$.  If a node has $k$ children, we assign $k-1$ inorder numbers to it.
(Unary nodes are not assigned any inorder number.)
If a node has more than one inorder number, we use the smallest value as its representative inorder number. 
Now we consider two operations (i) $irank(k, i)$: return the $i$-th inorder rank of S node $k$ (given as preorder number), and 
(ii) $iselect(j)$: given an inorder rank $j$ of an S node,
return $(k, i)$ where $k$ is the preorder number of the node with inorder rank $j$ and 
$i$ is the number such that 
$k$ is the $i$-th inorder number of the node. 
The following describes how to support both queries in $\Order(1)$ time using $\order(n)$ bits of additional space.

One can observe that for each micro-tree (or mini-tree) $t$ of $T_G$, all the inorder numbers corresponding to the S nodes in $t$ form two
intervals $I^1_t = [l^1_t, r^1_t]$ and $I^2_t = [l^2_t, r^2_t]$. Note that all the intervals corresponding to the mini-trees or micro-trees partition the interval $[1, \mathcal{S}]$ where $\mathcal{S}$ is the largest inorder number in $T_G$. 
We construct a dictionary $D_M$ that stores the right end points of all the intervals corresponding to the mini-trees, where with each element of the dictionary, we store a pointer to the mini-tree corresponding to that interval as the satellite information. The number of elements in this dictionary is $O(s/L)$ with universe size at most $s$, and hence can be represented as an FID~\cite{rrr} using $o(s)$ bits to support membership, rank and select queries in $\Order(1)$ time. The satellite information can also be stored in $o(s)$ bits, to support $\Order(1)$-time access.
For each mini-tree $M$, we also construct a dictionary $D^M_{\mu}$ that stores the right end points of all the intervals corresponding to its micro-trees, where with each element we associate a pointer to the corresponding micro-tree as the satellite information. The space usage of the dictionaries corresponding to all the mini-trees adds up to $o(s)$ bits in total.


In addition, for each mini-tree $T$ of $T_G$, we store its corresponding intervals $I^1_T$ and $I^2_T$ using $\order(s)$ bits in total. We call the two values $l^1_T$ and $l^2_T$ as the \textit{offsets} corresponding to $T$. Also, for each micro-tree $t$ contained in the mini-tree $T$ and $i \in \{1, 2\}$, we store $\{[l^i_t -l^1_T, r^i_t - l^1_T]\}$ if $I^i_t \subseteq I^1_T$, and $[l^i_t -l^2_T, r^i_t - l^2_T]$ otherwise (i.e., offsets with respect to the mini-tree intervals they belong to). Since all the endpoints of these intervals are at most $L$, we can store all such intervals using $\order(s)$ bits in total. 
The total space usage is $\order (s)$ bits. 

To compute $irank(k, i)$, we first find the micro-tree $t$ which contains the node $k$. 
Then, we decode the interval corresponding $t$ using the interval stored at $t$ as well as the offsets corresponding to $t$, and return the $i$-th smallest value within the interval. To compute $iselect(k)$, we first find micro-tree $t$ that contains the answer by the rank queries on $D_M$ and $D^M_{\mu}$. 
Finally, we compute the answer within the micro-tree $t$ in $\Order(1)$ time using the intervals stored with $t$.




Next, we assign labels to the vertices of the graph.
Any vertex in the graph corresponds to a common terminal of two SP graphs which are combined by series composition. For each vertex $v \in G$, let $S_v$ be an S node in $T_G$ which represents such series composition. Then we assign one inorder number of $S_v$ as the label of $v$ (note that any two subgraphs  which have a common terminal correspond to the subtree at the consecutive child nodes of $S_v$).
For example, vertex 5 in the graph corresponds to the common terminal of the following two subgraphs: (i) the subgraph consisting of the edge $g$ from 4 to 5, and (ii) the subgraph corresponding to the subtree rooted at the mini-tree $H$ (consisting of a single P node), which contains the four edges $h, j, i$ and $k$. 
Note that the inorder number 5 is assigned to the S node corresponding to the mini-tree $F$, when we traverse from subtree corresponding to (i) to the subtree corresponding to (ii) (during the preorder traversal of T).

Also, we define a label for each node $v$ of $T_G$, which is an ordered pair $(l_v,r_v)$ of the two terminals of the subgraph corresponding to the subtree rooted at that node.
We call $l_v$ and $r_v$ the left and the right label of the node $v$.
The label $(l_v, r_v)$ of a P node $v$ can be computed in $\Order(1)$ time as follows.
(1) If $v$ is the leftmost child of its parent $p$, 
then $r_v$ is equal to the smallest inorder number of $p$, 
given when $p$ is visited from $v$.
To obtain $l_v$, we traverse the SP tree $T_G$ up from $v$ until we reach an S node $q$ such that $v$ does not belong to the leftmost subtree of $q$.
We can compute the node $q$ in $\Order(1)$ time as follows.
If $q$ is in the same micro-tree as $v$, then we can find $q$ using a table lookup.
Otherwise, if $q$ is in the same mini-tree as $v$, then we store $q$ with the root of the micro-tree containing $v$.
Finally, if $q$ is not in the same mini-tree, then we explicitly store $q$ with the root of the mini-tree containing $v$.
%
(2) If $v$ is the rightmost child of its parent $p$, 
then $l_v$ is equal to the inorder number of $p$, 
given when $p$ is visited the last time before visiting $v$.
To obtain $r_v$, we traverse the SP tree $T_G$ up from $v$
until we reach an S node $p$ such that $v$ does not belong to the rightmost subtree of $p$.
We use a similar data structure as in (1) to compute the answer.
(3) In all other cases, $l_v$ and $r_v$ are the inorder numbers
of the parent $p$ of $v$, defined immediately before visiting $v$ from $p$, 
and immediately after visiting the next sibling of $v$ from $p$, respectively.

The label of an S node is the same as its parent P node (we don't assign a label to the root S node).
The label of a leaf can be determined by the same algorithm
for P or S nodes depending on whether its parent is an S or P node.
Note that, from the above definition, the label of a P node is the same as the label of any of its child S nodes.
For an S node $v$, suppose $v_1, v_2,\ldots, v_k$ be its $k$ children,
and $(\ell_1, r_1), (\ell_2, r_2), \ldots, (\ell_k, r_k)$ be the left and the right labels.
Then it holds that $r_1 = \ell_2, r_2 = \ell_3, \ldots, r_{k-1} = \ell_k$,
and the label of $v$ is $(\ell_1, r_k)$.

We also define $b(u)$ and $f(u)$ for each vertex $u$ of the graph, as follows.
Suppose that during the preorder
traversal of the tree, we visit nodes
$x$, $p$, $y$ in this order and we assign
the inorder number $u$ to $p$.
Then we define $b(u) = x$ and $f(u) = y$.
If $iselect(u)$ returns the pair $(p,j)$, then $x$ and $y$ are the $j$-th and the $(j+1)$-th children of node $p$, respectively. Thus, $b(u)$ and $f(u)$ can be computed in $\Order(1)$ time.
This completes the description for encoding of SP graphs. \\
%
%
\newline
\noindent\textbf{Supporting navigation queries.} 
For SP graphs, we additionally consider \emph{multiplicity} $(u, v)$ queries, which returns the number of edges between $u$ and $v$.
\begin{enumerate}
    \item $\bm{\adjacent{}(u,v)}:$
    Without loss of generality, assume that $u < v$. We first find 
    the nodes $b(u)$, $f(u)$, $b(v)$ and $f(v)$. 
    (1) If $f(u) = b(v)$, the subgraph corresponding to the node $f(u)$ has terminals with labels $u$ and $v$.
    Therefore $u$ and $v$ are adjacent if $f(u)$ is a leaf (this corresponds to the edge $(u, v)$)
    or $f(u)$ has a leaf child ($f(u)$ is a P node and it has a leaf child that corresponds to the edge $(u, v)$).
    (2) If $depth(b(u)) > depth(b(v))$, find the label of $f(u)$.
    Let $(u, x)$ be the label of $f(u)$.
    Then $u$ and $v$ are adjacent iff $x = v$, and $f(u)$ is either a leaf or is a P node with a leaf child.
    (3) If $depth(b(u)) < depth(b(v))$, 
    find the labels of $b(v)$.
    Let $(y, v)$ be the label of $b(v)$.
    Then $u$ and $v$ are adjacent iff $y = u$, and $b(v)$ is either a leaf or is a P node with a leaf child.    
    In all three cases, the query can be supported in $O(1)$ time.

\item $\bm{\multiplicity{}(u,v)}:$
 Again, without loss of generality, assume that $u < v$.
 If $\adjacent{}(u, v) = false$, then we return $0$. 
 If not, we consider the three cases above, and describe how to support the multiplicity query. 
 For Case (1), if $f(u)$ is a P node (otherwise, we return $1$), we can answer the query by returning the number of leaf children of $f(u)$ (note that this can be supported in $O(1)$ time using the tree covering of $T_G$). For Case (2), if $f(u)$ is a P node with label $(u, v)$ (if $f(u)$ is a leaf node, we return $1$), we answer the query by returning the number of leaf children of $f(u)$. Case (3) is analogous to Case~(2).

\item $\bm{\neighbor{}(u)}:$
    First we find $b(u)$ and $f(u)$.
    Then we apply the following procedure to explore all the neighbors of $u$ by executing the two procedure calls, ${\rm Explore}(b(u), \mbox{R})$ and ${\rm Explore}(f(u), \mbox{L})$.
    \begin{quote}
    ${\rm Explore}(x, \mbox{D})$: if $x$ is a leaf with label $(u,w)$ or $(w,u)$, then output~$w$.\\
     If $x$ is an S node, then call ${\rm Explore}(y, D)$, where $y$ is the leftmost (rightmost) child of $x$, if $D= \mbox{L}$ ($D= \mbox{R}$).
     If $x$ is a P node, then call ${\rm Explore}(y, D)$ for all the children $y$ of $x$.
    \end{quote}
 The running time of this procedure is proportional to the size of the output.
 Note that if we do not want to report the same neighbour multiple times, we can define a canonical ordering between the children of P nodes such that all the leaf children appear after the non-leaf children (S nodes), and only report the first leaf child of the node.
\item $\bm{\degree{}(u)}:$ 
    Let $\mu$ and $M$ be the micro-tree and mini-tree containing $u$ respectively. 
    Then the degree of $u$ is the summation of (i) the number of adjacent vertices in $\mu$, (ii) the number of adjacent vertices not in $\mu$ but in $M$, and (iii) the number of adjacent vertices not in $M$, denoted by $d^1_u$, $d^2_u$, and $d^3_u$ respectively. 
    Here an adjacent vertex of $u$ refers to a vertex $v$ such that $(u,v)$ or $(v,u)$ is the label of some leaf node.
    If $u$ is not one of the labels of the boundary node of $\mu$ (of $M$), then $d^2_u = 0$ (respectively, $d^3_u = 0$).
    Now we consider three cases as follows. First, 
    $d^1_u$ can be computed in $\Order(1)$ time using a precomputed table. 
    The value $d^2_u$ ($d^3_u$) can be stored with the root of micro-tree (mini-tree) whose parent is the boundary node in $\mu$ ($M$). 
    Note that in the above scheme, we only need to store two values corresponding to the two labels of the root, for each micro-tree/ mini-tree root. Thus the space usage for storing these values is $o(n)$ bits.

\end{enumerate}

\section{Block/Cactus/Block-Cactus Graphs}\label{sec:block}
A {\em block graph} (also known as a clique tree or a Husimi tree~\cite{doi:10.1063/1.1747725}) is an undirected graph in which every block (i.e., maximal biconnected component) is a clique. 
A {\em cactus graph} (same as {\em almost tree(1)}~\cite{GurevichSV84}) is a connected graph in which every two simple cycles have at most one vertex in common (equivalently every block is a cycle). A {\em block-cactus graph} is a graph in which every block is either a cycle or a complete graph.

\begin{figure}[htbp]
\begin{center}
\includegraphics[clip, width=12.0cm]{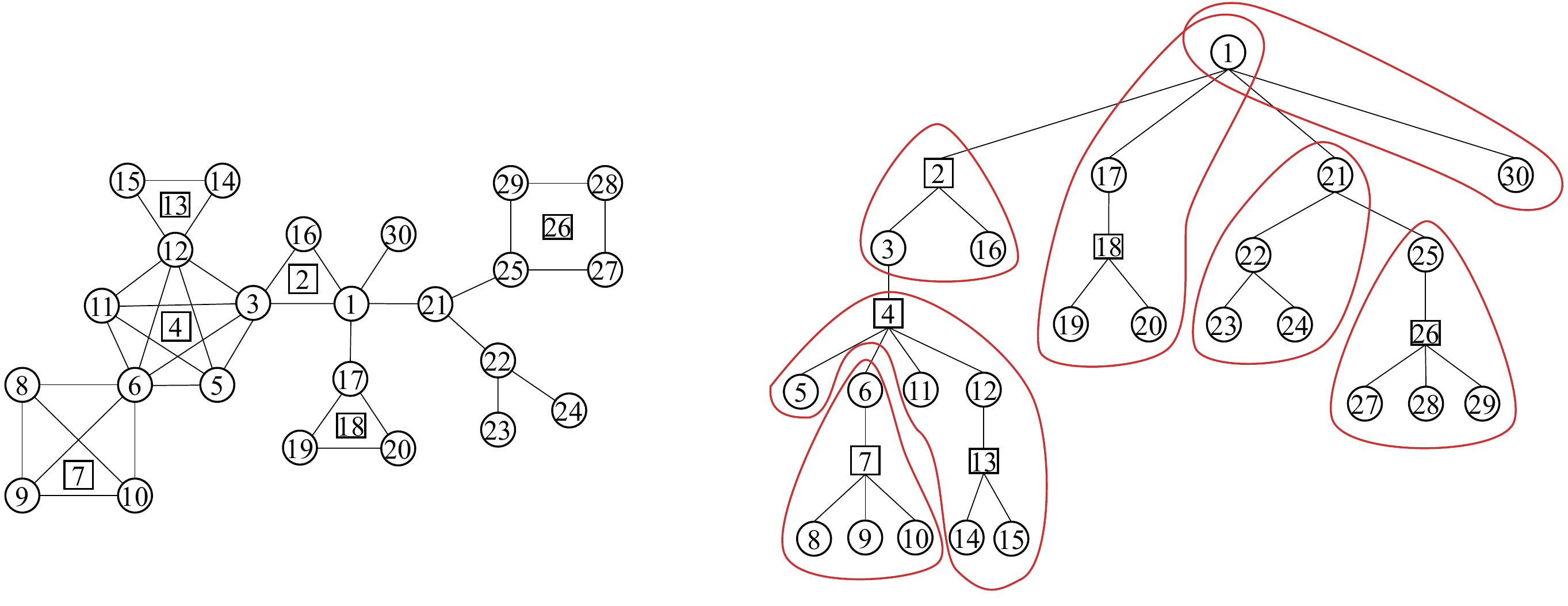}
\caption{An example of a block-cactus graph (left) and its tree representation (right).
Squares are dummy vertices.}
\label{fig:bc1}
\end{center}
\end{figure}

Any graph that belongs to one of these three graph classes can be converted into a tree as follows. Replace each block (either a clique or an induced cycle) with $k$ vertices by a star graph $K_{1,k}$ by introducing a dummy node that is connected to the $k$ nodes that correspond to the $k$ vertices of the block. The remaining edges and vertices of the graph are simply copied into the tree. See Figure~\ref{fig:bc1} for an example. 
Note that 
the number of dummy nodes is always less than the number of non-dummy nodes. 
In the following, we describe a succinct encoding for block-cactus graphs, and note that it is easy to obtain succinct encoding for block graph and cactus graph using these ideas.\\
\newline
\noindent\textbf{Succinct representation.}
Let $G$ be the input block-cactus graph, and let $T_G$ be the corresponding tree obtained by replacing each block with a star graph, as described above. We apply the tree covering algorithm of Theorem~\ref{th:originalTC} on $T_G$ with mini-tree and micro-trees of size $L = \log^2 n$ and 
$\ell = (\log n) / (2\alpha)$ for some constant $\alpha \ge 2.092$ 
respectively.

It is easy to see that each micro/mini-tree obtained by the tree cover algorithm corresponds to a block-cactus graph, although it may not be a subgraph of the original graph $G$. And by storing some additional information with each micro/mini-tree along with its representation, we can give a bijective map between the vertices in $G$ and the nodes in $T_G$, which we use in describing the query algorithms.

We first note that when we convert a block ($C_k$ or $K_k$) into a star graph ($K_{1,k}$), the neighbors of the dummy node can be ordered in multiple ways when we consider the resulting graph as an ordered tree. In particular, if the ordered tree is rooted at a dummy node corresponding to a cycle, then its children can be ordered in either the clockwise or anti-clockwise order of the cycle, and also the first child can be any vertex on the cycle. 
When the root of micro-tree $t$ is a dummy node corresponding to a cycle, the cycle corresponding to the dummy node
is cut into two or more pieces, and the one inside $t$ represents a shorter cycle.  Then the micro-tree $t$ is encoded as a canonical representation of the modified subgraph, and it loses the information of how it was connected to the other part of the graph.
To recover this information, for the micro-tree it is enough to store one vertex in the shorter cycle that is connected to the outside and the direction (clockwise or anti-clockwise) of the cycle.
The vertex is encoded in $\Order(\log\ell)$ bits, and the direction in one bit.
We need the same information for the non-root boundary node of the micro-tree.
This additional information will enable us to reconstruct the cycle in the original graph from the subgraphs corresponding to the micro-trees.
Note that if the dummy node corresponds to a clique, then we don't need this information.

Each micro-tree is encoded as a two-level pointer into a precomputed table that stores the set of all possible block-cactus graphs on at most $\ell$ vertices. 
Note that the number of dummy nodes is $\Order(n/\log n)$ since we can delete all the dummy nodes which are not boundary nodes of micro-trees.
We also store $1$ bit with each of these $\Order(n/\log n)$ dummy nodes, indicating whether it corresponds to a clique or a cycle. 
Thus each micro-tree is represented optimally, apart from an $O(\log \ell)$-bit additional information. Hence the overall space usage is succinct.
This completes the description for the succinct encoding of block-cactus graphs. \\
\newline
\noindent\textbf{Supporting navigation queries.} 
\begin{enumerate}
    \item $\bm{\adjacent{}(u, v)}:$ If there is an edge in $T_G$ between the nodes corresponding to $u$ and $v$, then $u$ and $v$ are adjacent in the graph (since we only delete some edges from the original graph; and all the edges added are incident to some dummy node). Otherwise, $u$ and $v$ are adjacent if they are connected to the same dummy node $x$, and either (a) $x$ corresponds to a clique, or (b) $u$ and $v$ are ``{\em adjacent}'' in the tree -- i.e., if they are adjacent siblings or one of them is the parent of $x$ and the other is either the first or last child of $x$. Since all these conditions can be checked in $\Order(1)$ time using the tree representation, we can support the query in $\Order(1)$ time.
    \item $\bm{\neighbor{}(u)}:$ The algorithm for this follows essentially from the conditions for checking adjacency. More specifically, to report $\neighbor{}(u)$, we first output all the non-dummy nodes adjacent to $u$ in the tree. And if $u$ is adjacent to any dummy node $x$, then we also output all the vertices: (a) that are connected to $x$ if $x$ corresponds to a clique, and (b) that are ``{\em adjacent}'' to it in the tree if $x$ corresponds to a cycle. This can be done in time proportional to the output size.
    \item $\bm{\degree{}(u)}:$ From the algorithm for the $\neighbor{}(u)$ query, we observe that the degree of a node can be computed by adding the two quantities: (1) the number of non-dummy neighbors of $u$, and (2) the number of nodes that are adjacent to $u$ through a dummy neighbor. It is easy to compute (1) and (2) within a micro-tree, in constant time using precomputed tables. In addition, we may need to add the contributions from outside the micro-tree, if $u$ is either a boundary node or is adjacent to a boundary node which is dummy. For each such dummy boundary node, we need to add either $1$ or $2$ (if the dummy node corresponds to a cycle) or $k$ (if the dummy node corresponds to a clique of size $k$). Since there are at most two such boundary nodes which can be adjacent to $u$, this can be computed in constant time. Also, for the roots of the mini (micro) trees, which are non-dummy, we store their degrees (within the mini-tree) explicitly. Thus, we can compute the $\degree{}(u)$ query in $O(1)$ time. 
\end{enumerate}

\section{$3$-Leaf Power Graphs}\label{sec:3leaf}
\begin{figure}[bt]
\begin{center}
\includegraphics[clip, width=11cm]{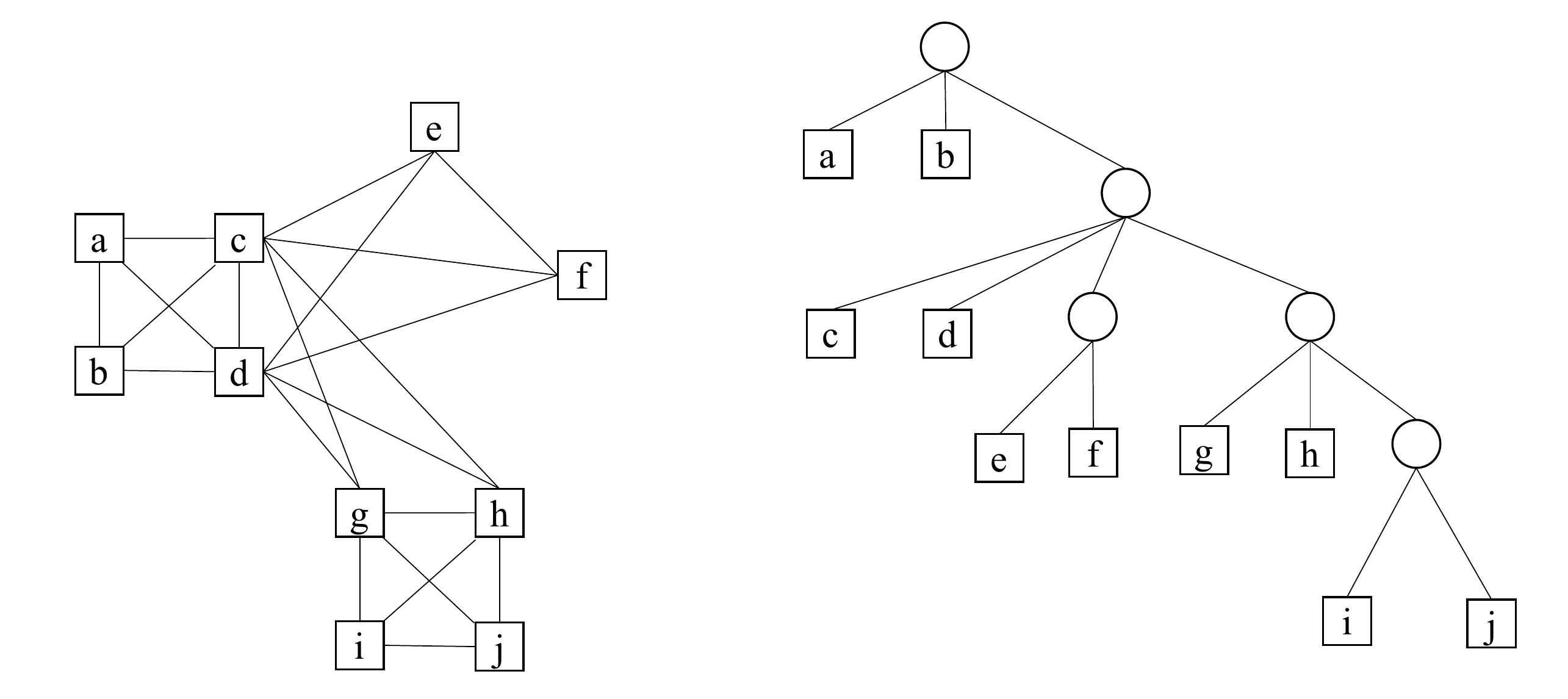}
\caption{An example of a $3$-leaf power graph (left) and its $3$-leaf root (right).}
\label{fig:3leafpower}
\end{center}
\end{figure}

A graph $G$ with $n$ vertices is a \textit{$k$-leaf power} if there exists a tree $T_G$ with $n$ leaves where each leaf node corresponds to a vertex in the graph $G$, and any two vertices in $G$ are adjacent if and only if the distance between their corresponding leaves in the tree is at most $k$. The tree $T_G$ is called a $k$-\textit{leaf root} of $G$ (see Figure~\ref{fig:3leafpower} for an example).
In this section, we consider the succinct representation of $k$-leaf power for the special case of $k = 3$. 
\\\\
\noindent\textbf{Succinct representation.}
Our representation of 3-leaf power graphs is based on the following lemma.

\begin{lemma}[Brandst{\"a}dt and Le~\cite{BrandstadtL06}]\label{lem:3-leafrecog}
For any connected and non-clique 3-leaf power $G$ of $n$ vertices, one can construct a unique 3-leaf root $T_G$ of $G$ of $O(n)$ nodes.
\end{lemma}

Note that We can make $T_G$ as a rooted tree as follows.
Because $T_G$ contains an internal node (otherwise $T_G$ consists of just an edge with two nodes, which corresponds 
to the clique $K_2$), we regard it as the root of $T_G$.
We store the root of every micro tree of $T$ explicitly using $\Order(n/\ell \cdot \log \ell) = \order(n)$ bits in total. 

Now consider the 3-leaf root $T_G$ of $G$. If $G$ is not a clique, one can construct the unique representation of $T_G$ by Lemma~\ref{lem:3-leafrecog}. If not, we fix $T_G$ as $K_{1, n}$. 
For any non-leaf node $p \in T_G$, we order the children of $p$ in the non-decreasing order of the sizes of the  subtrees rooted under them (thus, all the leaf children of $p$ appear before the non-leaf children of $p$), to support the navigation queries efficiently.
We then apply the tree covering algorithm on $T_G$ with parameters $L = \log^2 n$ and $\ell = (\log n )/(2\alpha)$ for any constant $\alpha \ge 1.35$. We build a precomputed table of size $\order(n)$-bits  which stores all non-isomorphic non-clique 3-leaf powers of size at most $\ell$ along with their 3-leaf root constructed from the algorithm of Lemma~\ref{lem:3-leafrecog}.

We use the following properties of 3-leaf roots:
(1) if $G$ is connected, every internal node of $T_G$ has at least one leaf child, and
(2) the graphs corresponding to the micro-trees created by applying the tree cover algorithm to $T_G$ 
are connected.
The proofs are as follows.  For (1), assume to the contrary that there is an internal node $v$ 
with no leaf children. Then any vertex corresponding to a leaf descendant of $v$ is not connected to any other vertex corresponding to a leaf node outside of the subtree rooted at $v$ since the distance between them (in $T_G$) is at least 4.
For (2), consider a micro-tree $t$ with a boundary edge connecting a node $v$ in $t$ and
a node $w$ which is the root of another micro-tree $t'$.  From (1), $v$ has a leaf child $u$.
If $u$ belongs to $t$, the graph corresponding to $t$ is connected.
If $u$ belongs to $t'$, the root of $t'$ must be $v$, which contradicts the assumption
that $(v,w)$ is a boundary edge.
Note that root boundary edges do not effect the connectivity of the graph corresponding to the micro-tree.

Thus each micro-tree $t$ of $T_G$ falls into one of the three cases: (i) 3-leaf root of a non-clique, (ii) single non-leaf node, or (iii) 3-leaf root of a clique. 
For Case (i), we encode $t$ as an index into the precomputed table. 
For Case (ii), we add one extra entry into the precomputed table, which is used to encode this case. 
Finally for Case (iii), note that there are only $\ell$-distinct 3-leaf roots corresponding to the clique of size $(\ell-1)$, each of which can be constructed by connecting two non-leaf nodes of $K_{1, i-1}$ and $K_{1, (\ell-i)}$ for any $1 \le i \le \ell-1$ (assuming $K_{1, 0}$ corresponds to the empty graph).
Thus, we add $\Theta(\ell^2)$ extra entries into the precomputed table which indicate cliques of size at most $\ell$ with an additional index $0 \le i \le \ell$. Overall, the total space of the encoding is succinct.
\\

\noindent\textbf{Supporting navigation queries.} 
For the navigation queries, we refer to each vertex $u \in G$ by the leaf-rank of the corresponding leaf node $l_u$ in $T_G$. Let $p_u$ be the parent node of $l_u$, and let $C_u$ and $D_u$ be a set of leaf and non-leaf children of $p_u$ respectively.
\begin{enumerate}
\item $\bm{\adjacent{}(u, v)}:$ 
By the definition of 3-leaf root, $l_u$ and $l_v$ are adjacent if and only if (i) $p_v = p_u$ or (ii) $p_u$ is a parent node of $p_v$ or vice versa. Since both $p_u$ and $p_v$ can be computed in $\Order(1)$ time~\cite{Farzan2014}, we can answer the $\adjacent{}(u, v)$ in $\Order(1)$ time.
\item $\bm{\neighbor{}(u)}:$
Let $s_u$ be a parent node of $p_u$. Then $j \in \neighbor{}(u)$ if and only if $l_j$ is a (i) leaf child node of $s_u$, (ii) node in $C_u$, or (iii) leaf child node of the node in $D_u$.  
To return all the leaf children of $s_u$, we scan from the leftmost child of $s_u$, and return if the child is a leaf node. This can be done in $\Order(1)$ time per node by using the $\Order(1)$-time tree navigation queries in \cite{Farzan2014}.
Next, we scan all the children of $p_u$. While scanning the node $u'$, if  
$u' \in C_u$ (this returns all the nodes in the case (ii)), we return $u'$. Otherwise, we return all the leaf-children of $u'$ (this returns all the nodes in the case (iii)).  
Again, all these nodes can be reported in $\Order(1)$ per node by the same argument as the above. 
Thus, we can return $\neighbor{}(u)$ in $O(|\degree{}(u)|)$ time.
\item $\bm{\degree{}(u)}:$ We count the number of (i) leaf child nodes of $s_u$ (parent node of $p_u$), (ii) nodes in $C_u$, and (iii) leaf children of the nodes in $D_u$ separately, and return the sum of these as the answer of $\degree{}(u)$ query. Now we describe how to compute (iii) in $\Order(1)$ time (note that (i) and (ii) also can be computed in $\Order(1)$ time analogously). Let $t_1$ (resp. $t_2$) be a micro-tree (resp. mini-tree) which contains $l_u$. Then we first consider the case that $D_u$ does not contain the boundary node of $t_1$. 
In this case, we compute the (iii) in $\Order(1)$ time using the precomputed table if $p_u$ is not a boundary node of $t_1$. If $p_u$ is a boundary node of $t_1$ (resp. $t_2$), we compute the (iii) in $\Order(1)$ time by referring to the answer stored at the root of $t_1$ (resp. $t_2$). Note that we can store all of these answers using at most $\Order(n/L \cdot \log L + n / \ell \cdot \log \ell ) = \order(n)$ bits in total. Next, we consider the case that $D_u$ contains the boundary node of $t_1$. In this case, we additionally store the number of leaf children of the root node of each micro-tree of $T_G$ using at most $o(n)$ bits in total. Then we can compute the (iii) in $\Order(1)$ time by computing the (iii) without the number leaf children of the boundary node of $t_1$, and adding the number of leaf children of the micro-tree whose root node is the child of the boundary node of $t_1$.
\end{enumerate}



\section{Conclusions}
We present in this work succinct representations of 
series-parallel, block-cactus and 3-leaf power graphs along with supporting basic navigational queries optimally. 
We conclude with some possible future directions for further exploration. Following the works of~\cite{DBLP:conf/wads/AcanCJS19,FarzanK14}, is it possible to support shortest path queries efficiently on these graphs while using same space as in this paper? Is it possible to design space-efficient algorithms for various combinatorial problems for these graphs? 
Can we generalize the data structure of Section~\ref{sec:3leaf} to construct a succinct representation of $k$-leaf power graphs? 
Finally, can we prove a lower bound between the query time and the extra space i.e., redundancy, for our data structures? 

\bibliographystyle{splncs04}
\bibliography{ref}


\end{document}